# MgCNi$_3$: a conventional and yet puzzling superconductor

J.-Y. Lin[1] and H. D. Yang[2]

[1]*Institute of Physics, National Chiao Tung University, Hsinchu 300, Taiwan ROC*

[2]*Department of Physics, National Sun Yat-Sen University, Kaohsiung 804, Taiwan ROC*

The newly discovered superconductivity in MgCNi$_3$, though with $T_c \leq 8$ K lower than that of the celebrated MgB$_2$, is probably even more interesting in its many puzzling physical properties. MgCNi$_3$ has been theoretically speculated to be unstable towards ferromagnetism. However, there are numerous evidences from the specific heat $C(T)$, Tunneling spectroscopy and NMR experiments indicating conventional $s$-wave superconductivity in MgCNi$_3$. The Hall effect and the thermoelectric power experiments suggest that the carriers responsible for the transport properties are electrons, in obvious contrast to holes predicated by the band structure calculations. In this article, we report the results of $C(T)$ experiments, upper critical field $H_{c2}$ measurements, and the pressure effects on MgCNi$_3$. These experimental evidences clearly demonstrate that superconductivity in MgCNi$_3$ is well explained within the conventional electron-phonon interaction scenario, at most with minor modifications from the magnetic interaction. The thermodynamic data $C(T)$ is consistent with the conventional s-wave order parameter. $H_{c2}$ of all samples follows a universal WHH relation $H_{c2} \approx aT_c(dH_{c2}/dT)_{T_c}$. Surprisingly, $dT_c/dP$ is positive which leaves room for further improvements in band structure calculations. There are other serious discrepancies between experiments and theory like in the transport properties and x-ray photoemission. The possible reconciliation within the two-band model and the consequent difficulties are discussed.

## I. Introduction

Superconductivity in MgCNi$_3$ ($T_c \leq 8$ K) was discovered soon after that of MgB$_2$ ($T_c$=39 K) [1,2]. Though with $T_c$ lower than that of the celebrated MgB$_2$, MgCNi$_3$ is probably even more interesting in its many puzzling physical properties. The issues in MgCNi$_3$ are rather appealing. Being a perovskite superconductor like Ba$_{1-x}$K$_x$BiO$_3$ and cuprate superconductors, MgCNi$_3$ is special in that it is neither an oxide nor does it contain any copper. Meanwhile, very recently, several new superconductors have been reported with possible ferromagnetic spin fluctuations or with the coexistence of ferromagnetism and superconductivity [3,4]. These intriguing findings have stimulated much new interest in the interplay between ferromagnetism and superconductivity. MgCNi$_3$ can be regarded as fcc Ni with only one quarter of Ni replaced by Mg and with C sitting on the octahedral sites. With the structure so similar to that of ferromagnetic Ni, the occurrence of superconductivity in MgCNi$_3$ is really surprising. Actually, there has been a theoretical prediction that MgCNi$_3$ is unstable to ferromagnetism upon doping with 12% Na or Li [5]. In this context, MgCNi$_3$ could be a superconductor near the ferromagnetic quantum critical point [6,7]. A possible magnetic coupling strength due to spin fluctuations was proposed [8]. Even more, a *p*-wave pairing in MgCNi$_3$ was suggested to be compatible with the strong ferromagnetic spin fluctuations [5]. If it were a *p*-wave superconductor, it would be the one with highest $T_c$ (e.g., compared to Sr$_2$RuO$_4$ with $T_c \leq 1.5$ K).

To critically examine these interesting scenarios, a good number of research results have accumulated. These activities covered experiments of the electrical transport, specific heat, pressure effects, tunneling spectroscopy, photoemission, x-ray absorption, NMR, flux pinning and more. The studies on the theoretical side are also intensive to elucidate the puzzles brought by MgCNi$_3$. Some of these will be discussed in this article. As one shall see, there are a few inconsistent experimental results from different reports. More noticeably, at this moment, there exist discrepancies between theory and experiments on certain fundamental properties of MgCNi$_3$. In this article, we would establish several fundamental properties of MgCNi$_3$, show the inconsistencies between theory and experiments, and discuss the possibilities and difficulties of reconciliation.

## II. Experiments

The MgCNi$_3$ samples were prepared based on the procedure described in [1]. The starting materials were magnesium powder, glass carbon, and nickel fine powder. The raw materials were thoroughly mixed, then palletized and wrapped with Ta foil before sealed into an evacuated quartz tube. The sample was first sintered at 600$^\circ$C for a short time and ground before further treated in a similar way at 900$^\circ$C for 3 hours. The x-ray diffraction pattern revealed the nearly single phase of MgCNi$_3$ structure. Details of the sample preparation and characterization are described elsewhere [9].

Resistivity $\rho(T)$ was measured by the four probe method. Thermoelectric power ($S$) measurements were performed with the steady state techniques. $C(T)$ was measured using a $^3$He thermal relaxation calorimeter from 0.6 to 10 K with magnetic fields $H$ up to 8 T. Detailed description of the measurements can be found in [10]. The hydrostatic pressure ($P$) dependent ac magnetic susceptibility ($\chi_{ac}$) data were taken by the piston cylinder self-clamped technique [11]. The hydrostatic pressure environment around the sample was generated inside a Teflon cell with 3M Fluroinert FC-77 as the pressure transmitting medium. The pressure was determined by using Sn manometer situated near the sample in the same Teflon cell. In each instance, the original value was reproduced within experimental error after the pressure released indicating complete reversibility of the pressure effect.

Fig. 1 show ...($T$) and $S$($T$) of one of our typical samples. Both ...($T$) and $S$($T$) are consistent with those in the literatures [1,12,13]. In our experiments, $T_c$ of the samples may slightly vary due to various carbon contents [14]. Nevertheless, all samples are very similar in the magnitude and temperature dependence of ...($T$).

### III. Fundamental parameters derived from the low-temperature specific heat $C(T)$

For the MgCNi$_3$ sample used in $C(T)$ experiments, ...=217 and 93 $\mu\Omega$ cm at $T$=300 and 10 K, respectively. Magnetization, specific heat, and resistivity measurements all showed a superconducting onset at about 7 K in the present sample. The resistivity transition width is 0.5 K, while thermodynamic $T_c$ determined from $C(T)$ is 6.4 K (see below).

$C(T)$ of MgCNi$_3$ with $H$=0 to 8 T is shown in Fig. 2 as $C/T$ vs. $T^2$ [15]. The superconducting anomaly at $H$=0 is much sharper than that in Ref. [1], indicating high quality of the sample, and clearly persists even with $H$ up to 8 T. It is noted that $C/T$ shows an upturn at very low temperatures. This upturn disappears in high $H$, which is a manifestation of the paramagnetic contribution like the Schottky anomaly. The normal state $C_n(T)=\chi_n T+C_{lattice}(T)$ was extracted from $H$=8 T data between 4 and 10 K by $C(T, H$=8T$)=\chi_n T+C_{lattice}(T)+ nC_{Schottky}(g\sim H/k_BT)$, where the third term is a 2-level Schottky anomaly. $C_{lattice}(T)=sT^3+uT^5$ represents the phonon contribution. It is found that $\chi_n$=33.6 mJ/mol K$^2$. This value of $\chi_n$, with the electron-phonon coupling constant $\}$ estimated below, requires a higher band $N(E_F)$ than most of those reported from calculations [5,8,16-19]. $\Theta_D$ derived from $C_{lattice}$ is 287 K, impressively lower than that (450 K) of Ni. This low $\Theta_D$, nevertheless, is close to the estimate based on the softening of the Ni lattice [16], which could enhance the electron-phonon interaction. The concentration of paramagnetic centers can be estimated to be the order of 10$^{-3}$, consistent with the estimate from the magnitude of the low temperature upturn. With a dominant content of Ni in this compound, this number is understandable and the paramagnetic contribution was indeed observed in the magnetization measurements, too [9].

To elucidate superconductivity in MgCNi$_3$, it is of interest to derive $\Delta C(T)=C(T)-C_{lattice}(T)-\chi_n T$. The resultant $\Delta C(T)/T$ at $H$=0 is shown in Fig. 3(a). By the conservation of entropy around the transition, the dimensionless specific jump at $T_c$ $\Delta C/\gamma_n T_c$=1.97±0.10 as shown in Fig. 3(b). This value of $\Delta C/\gamma_n T_c$ is very close to that in [1], though with a sharper transition in the present work. If the relation of $\Delta C/\gamma_n T_c$=(1.43+0.942$\lambda^2$-0.195$\lambda^3$) [20] is adapted as was in Ref. [1], $\lambda$ is estimated to be 0.83. Both values of $\Delta C/\gamma_n T_c$ and $\lambda$ suggest that MgCNi$_3$ is a moderate-coupling superconductor rather than weak-coupling. To compare $\Delta C(T)$ of MgCNi$_3$ with a BCS one, $\Delta C(T)/T$ from the BCS model with $2\Delta/kT_c$=4 was plotted as the solid line in Fig. 3(a). There was no attempt to fit data with the BCS model. The choice of $2\Delta/kT_c$=4 instead of the weak-coupling value 3.53 was somewhat arbitrary and was to account for the larger $\Delta C/\gamma_n T_c$=1.97 than the weak-limit one 1.43. However, it is noted that already the data can be well described by the solid line at least qualitatively, except the low temperature part of data which suffer contamination from the magnetic contribution. With this very magnetic contribution, it is difficult to check the thermodynamic consistency. Nevertheless, if the data below 3 K are replaced by the solid line, entropy is conserved as shown in the inset of Fig. 3(a). It is worth noting that $\Delta C(T)/T$ of MgCNi$_3$ is *qualitatively* different from that of Sr$_2$RuO$_4$, which is considered as a $p$-wave superconductor [21].

To further examine $C_{es} \equiv C(T,H)-C_{lattice}(T)$, $C_{es}(T)/\gamma_n T_c$ vs. $T_c/T$ for $H$=0 was plotted in Fig. 4. The fit of data between 2 and 4.5 K leads to $C_{es}/\gamma_n T_c$=7.96exp(-1.46$T_c/T$). Both the values of the prefactor and the coefficient in the exponent are typical for BCS superconductors. Since the magnetic contribution would make $C_{es}$ overestimated at low temperatures, the value of 1.46 in the exponent is probably slightly underestimated. This is in contrast to the case of MgB$_2$, in which $C_{es} \propto$ exp(-0.38$T_c/T$) [10,22]. This small coefficient in the exponent for MgB$_2$ is usually attributed to a multi-gap order parameter.

For a gapped superconductor, $\chi(H)$ is expected to be proportional to $H$ [23]. For nodal superconductivity, $\chi(H) \propto H^{1/2}$ is predicted [24]. Actually, $\chi(H)$ of cuprate superconductors has been intensively studied in this context [25]. To try to figure out $\chi(H)$ in MgCNi$_3$, $\chi(H)$ vs. $H$ and $\delta C(T,H)/T (\equiv C(T,H)/T-C(T,0)/T)$ vs. $H$ at 2 K is shown in Fig. 5(a) and (b), respectively. In Fig. 5(a), $\chi(H)$ is derived from the extrapolation of the data in Fig. 2 to $T$=0 [26]. Since the magnetic contribution is rather significant for low field data at 0.6 K, only $\chi(H)$ with $H \geq$ 4 T can be derived in this way. Clearly, $\chi(H)$ follow a straight line passing through the origin, which suggests $\delta \chi \propto H$. At $T$=2 K, the magnetic contribution is not so significant as at 0.6 K, thus $C_{es}(T,H) \approx C_{es}(T,H=0)+\chi(H)T$ [27,28]. This approximation neglects the temperature dependence of $\Delta$. However, since $\Delta$ varies slowly below $T_c/2$, information of $\chi(H)$ can still be deduced this way reliably as one shall see from $d\chi/dH$. $\delta C/T$ in all magnetic fields are shown as the solid circles. As seen in Fig. 5(b), all high field data can be well described by the straight line, indicating again a linear $H$ dependence of $\gamma$. Data below $H$=1 T begin to deviate from the linear behavior due to flux line interactions at low $H$, as nicely demonstrated in [27]. The straight line passes through the origin in Fig 5(a), which implies that the flux line interactions are relatively insignificant compared to the core contribution at very low temperatures. This trend was also observed in [27]. The slopes $d\gamma/dH$ in Fig. 5(a) and (b) are 2.91±0.05 and 3.15±0.08 mJ/mol K$^2$ T, respectively. These two close values derived from different methods suggest that the relation $\delta\gamma \propto H$ is genuine. On the other hand, one could try to fit the data in Fig. 5(b) by $\delta\gamma(H) \propto H^{1/2}$. The results are represented by the dashed line in Fig. 5(b). Apparently, the data can not be well described in this manner, in contrast to the nice $\delta\gamma(H) \propto H^{1/2}$ relation found in cuprates [29-33]. In principle, it is possible to obtain $\gamma(H)$ in low fields by subtracting the paramagnetic contribution $C_m(T,H)$ from $C(T,H)$. There were attempts to obtain $\gamma(H)$ by taking $C_m(T,H)$ the form of the Schottky anomaly [15,26]. It was found the derived $\gamma(H)$ could be different depending on the details of the fitting. This implies that the Schottky anomaly can not totally account for the magnetic contribution at low temperatures. The results in [15,26] are intriguing but inconclusive.

## IV. The upper critical field $H_{c2}$

The upper critical field $H_{c2}$ is one of the important fundamental properties of a type II superconductor. It is related to important parameters like the coherence length $\xi$ and many magnetic properties. The value of $H_{c2}$ is also one of the crucial criteria for applications of the superconducting magnets. $H_{c2}$ can also be deduced from the magnetic field dependence of $C$. It is of interest to see if the values of $H_{c2}$ by distinct techniques are consistent with each other. In the previous section, the magnetic field dependence of the linear coefficient $\gamma$ in $C$ from the electronic contribution is proportional to $H$ in MgCNi$_3$, and this can be taken as one of the strong evidences for $s$-wave pairing [15]. This conclusion in principle could be further examined by the relevant data of $H_{c2}$. In this section, both $C$ and $\rho$ were measured *using the same sample*. This would allow leading to more conclusive results by eliminating any possible uncertainty from individual samples.

Fig. 6 shows the transition of $\rho(T)$ in $H$ up to 8 T for the same sample used in $C(T)$ measurements [15]. The normal state $\rho(T)$ of the present sample is slightly smaller than that in [12]. It is noted that $H$ did not broaden the transition width, probably due to the low temperatures investigated here. $T_c$ is determined by the criterion of 50% of the normal state $\rho$ just above $T_c$.

$H_{c2}$ vs. $T$ resulting from both $C$ and $\rho$ measurements is shown in Fig. 7 [34]. $T_c$

determined by $C$ is slightly lower than that by $\chi$ as expected. Other than that, both techniques yield very similar results. $H=$ 8T leads to about 50% of $T_c$ suppression. The data show a downward curvature in Fig. 7 as in conventional superconductors. The slopes $(dH_{c2}/dT)_{T_c}$ derived from the linear fit of both the $C$ and $\chi$ data for $0.8\leq(T/T_c)<1$ are very close to each other. The values of $(dH_{c2}/dT)_{T_c}$ are 2.96±0.08 T/K and 2.88±0.03 T/K from $C$ and $\chi$ measurements, respectively. The relation of $H_{c2}\approx 0.69 T_c(dH_{c2}/dT)_{T_c}$, which does not take into account the effects of the spin-orbit interaction and the spin paramagnetic term [35], was used in Ref. [12] To estimate $H_{c2}$ at $T=0$. In this way, it would lead to $H_{c2}$=13.2±0.7 T in the present sample. However, fail to include the spin-orbit and the spin paramagnetic effects is known to significantly overestimate $H_{c2}$. To derive $H_{c2}$ accurately, one has to utilize the numerical analysis which is beyond the scope of this article. However, certain plausible assumption can be made. It is found that the physical properties of MgCNi$_3$ are very similar to those of Nb$_{0.5}$Ti$_{0.5}$ [15]. It seems plausible for both compound to have a similar relation between $H_{c2}$ and $(dH_{c2}/dT)_{T_c}$. According to Ref. [35], $H_{c2}\approx 0.59T_c(dH_{c2}/dT)_{T_c}$ for Ni$_{0.5}$Ti$_{0.5}$. Following this relation, $H_{c2}$=11.2±0.6 T in the present sample. It is intriguing to derive $H_{c2}$ from $\chi(H)$. The high field $\chi(H)$ can be estimated from the linear extrapolation to $T=0$, and is found to be proportional to $H$ with $d\chi/dH$=2.91 mJ/mol K$^2$ T. For $\chi_n$=33.6 mJ/mol K$^2$, $H_{c2}$=11.5±0.6 T can be estimated, amazingly close to the above value estimated from the $T_c$ transition in $H$. Estimates of $H_{c2}$ for the present sample are summarized in Table I.

$H_{c2}$ of MgCNi$_3$ was reported in several other works mainly by resistivity $\chi$ measurements [12,19,36]. It is of interest to test if all the related $(dH_{c2}/dT)_{T_c}$ data in the literatures can be described ina universal way. In Fig. 8, data from four different reports are shown as $H_{c2}$ vs. $(dH_{c2}/dT)_{T_c}T_c$. The results suggest that all data follow the WHH formula and can be described by $H_{c2}\approx aT_c(dH_{c2}/dT)_{T_c}$ with a universal $a$. From the best fit of data in Fig. 8, $a$=0.60 consistent with the above postulation. This analysis resolves the difficulty in a seemingly too high $H_{c2}$ (compared to the one derived from the extrapolation to $T=0$ or from other methods) in the works using $H_{c2}\approx 0.69T_c(dH_{c2}/dT)_{T_c}$. The suggestion from Fig. 8 is also in contrast to a clean limit scenario of $H_{c2} \propto T_c^2$ proposed in Ref. [19], which mistakenly assumed a constant $(dH_{c2}/dT)_{T_c}$ for all samples and took too large values of $H_c$ as stated above.

To summarize, we have estimated $H_{c2}$ from the $T_c$ transition in $H$ by $C$ and $\rho$ measurements. Both lead to identical value of $H_{c2}$. Furthermore, this value is consistent with that derived from linear $\chi(H)$ with respect to $H$. Results in this section demonstrate that $H_{c2}$ of MgCNi$_3$ can be describe by WHH formula within the context of conventional superconductivity. Moreover, since $\chi(H) \propto H$ is characteristic of conventional superconductors, the results presented in this section have intriguing implications on the order parameter of MgCNi$_3$ discussed later.

## V. The pressure effects

It is well known that the high pressure ($P$) plays an important role on the $T_c$ of the intermetallic superconductors [37-42]. In general, $P$ can change the electronic structure, phonon frequencies or electron-phonon coupling that affecting $T_c$. Both positive and negative pressure derivatives, $dT_c/dP$, are observed in the metallic and intermetallic superconductors. For example, simple $s,p$-metal superconductors [43] like Al, In, Sn, or Pb, and the intermetallic superconductors like the recently discovered MgB$_2$ have shown the decrease in $T_c$ with the increase in $P$ [38-40]. However, depending on the rare earth site of the quaternary borocarbides, RNi$_2$B$_2$C (R= rare earths), both increase and decrease in $T_c$ were observed with the increase in pressure [41,42]. In addition, the pressure can basically shift the Fermi level (E$_F$) towards higher energies [41,42] and thereby provide a probe on the slope of the DOS near E$_F$. Moreover, it can also modify the magnetic pair breaking effect and tune the competitive phenomena between superconductivity and spin fluctuations. From the specific heat and magnetic field dependent resistivity studies of the previous two sections, it has been suggested that the MgCNi$_3$ is a typical BCS-like superconductor the moderate coupling. In this section, we further present the pressure effects on the $T_c$ of this unusual superconductor to investigate its unique electronic properties.

Three samples A, B, and C with different $T_c$'s due to various carbon contents, as showns in the inset of Fig. 1, were used in the pressure effect experiments. $T_c$ (midpoint) for sample A increases from 6.56 to 6.79 K with the increase in pressure from ambient to 14.80 kbar as shown in Fig. 10 with the rate of $dT_c/dP \sim 0.015$ K/kbar. The similar trend of pressure effect on $T_c$ for samples B and C is also shown in Fig. 9. The magnitude of $dT_c/dP$ for electron-carrier MgC$_x$Ni$_3$ is about the same as those of its two dimensional analogue $R$Ni$_2$B$_2$C ($R$= rare earths) superconductors though most of those are negative, except LuNi$_2$B$_2$C in which T$_C$ increases with pressure as in the present samples. On the other hand, the $dT_c/dP$ for hole-carrier MgB$_2$ is of one order of magnitude higher and is negative.

Assuming that the Coulomb interaction and the electron-phonon interaction is less pressure dependent, the change of $T_c$ with the unit cell volume ($V$) can be given by [37,41]

$$(V/T_c)(dT_c/dV) = d\ln T_c/d\ln V = - (B/T_c)(dT_c/dP) = g(d\check{S}/dV, d\check{S}/dV) \quad (1)$$

, where $B$ is the bulk modulus of the superconductor, $\check{S}$ is the phonon frequency and $g$ is a universal function. Using the calculated value of $B$ for MgC$_x$Ni$_3$ as 1510 kbar [44] and taking the $dT_c/dP$ and $T_c$ from Fig. 9, the values of $d\ln T_c/d\ln V$ are found from Eq. (1) respectively for samples A, B and C as –3.18, -2.58, and –2.76. These values are of the same order of magnitude in MgB$_2$ superconductor (+ 4.16) with opposite sign [37].

The positive value of $dT_c/dP$ in MgCNi$_3$ is somewhat surprising. $P$ usually shift $E_F$ toward higher energy due to the decrease in $V$ when $P$ in applied on the sample. If the positive value of $dT_c/dP$ is ascribed to an increase in $N(E_F)$, it would implies a positive slope of $N(E)$. On the other hand, all band structure calculations indicate the existence of von Hove sinfularity (vHs) lying below $E_F$ by ~0.1 eV. Naively, $P$ should have led to a negative value of $dT_c/dP$. The apparent discrepancy between experimental and theoretical results could be reconciled with each other within the two-band model proposed in [18,19]. This model consists of one almost full heavy (or dirty) hole band and one almost empty light (or clean) electron band that cross the Fermi energy. Although the large proportion of $N(E_F)$ comes from the hole band with vHs, superconductivity of MgCNi$_3$ is likely due to the electron carriers. Since $N(E)$ of the electron band has a positive slope at $E_F$, the positive value of $dT_c/dP$ is then explained.

Even though the strong spin fluctuations are unfavorable to exist in MgCNi$_3$ (see the following section), the marginal or unstable spin fluctuations suppressing $T_c$ have not been totally ruled out. In general, the pressure reduces the spin fluctuations. Because the spin fluctuations and superconductivity are mutually competitive phenomena, this may provide an alternative explanation for the positive pressure effect on the $T_c$ of MgC$_x$Ni$_3$.

Ref. [18] previously reported an initial decrease in $T_c$ followed by an increase on application of pressure. This curious $P$ dependence of $T_c$ is likely due to individual sample quality and the $T_c$ measurement method. It is noted that ... of the sample in [18] is one order of magnitude larger than the present samples, and $T_c$ in [18] was determined by ... rather than $t_{ac}$ in the present work [45]. Generally, $t_{ac}$ measurements probe the bulk properties of the sample, and ... measurements on polycrystalline samples are sometimes sensitive to the grain boundary. It has been shown that the grain boundary in MgCNi$_3$ is important to physical properties like flux dynamics [46]. There is a possibility that $T_c$ determined by $\omega$ also partially manifests the properties of the grain boundary. It is not implausible that the grains of the sample in [18] were weakly coupled and the initial application of $P$ only made the grain boundary more compact rather than reduced $V$ of the grains. Only with the strongly coupled grains, did the intrinsic pressure effects begin to show up and $T_c$ increased. Actually, $dT_c/dP$ was of the same order as that of the present work when $T_c$ increased with larger $P$ in [18]. These results suggest that experiments in both [18] and [45] measured the same pressure effects as long as $dT_c/dP$ was positive. The above scenario is further supported by the observation that $\omega$ of the sample in [18] decreased by almost 75% on the initial application of $P$.

## VI. Discussions

### A. The superconducting order parameter of MgCNi$_3$

Due to the proximity of ferromagnetism, superconducting order parameter in MgCNi$_3$ was expected to be *p*-wave by [5] and others. However, it is noted that the *s*-wave superconductivity in weak ferromagnetism phase was once proposed [6]. Since there is no evidence for nodal lines of order parameter from the specific heat data, nature must have chosen the gapped order parameter like $x+iy$ if it was *p*-wave in MgCNi$_3$. It is instructive to compare the physical parameters of MgCNi$_3$ with those of Nb$_{0.5}$Ti$_{0.5}$ and Nb, which are two s-wave superconductors. The results are listed in Table II. It seems that MgCNi$_3$ appears ordinary among these superconductors. $H_{c2}$ of Nb is much smaller than those of the others because Nb$_{0.5}$Ti$_{0.5}$ and MgCNi$_3$ are typical type II superconductors while Nb is nearly type II. (The coherence length $\xi \approx 5.6$ nm in the present MgCNi$_3$ sample, and the preliminary magnetization measurements suggest a penetration depth $\lambda_L$=128-180 nm [9].)

It can not be overemphasized that *not only does the strong evidence of the conventional s-wave superconductivity in MgCNi$_3$ come from the exponential behavior of C(T) in Fig. 4, but also as importantly from the well thermodynamic description of s-wave superconductivity in Fig. 3(a) and the comparative studies in Table II.* Furthermore, results from both $C(T)$ and $H_{c2}$ experiments are consistent with each other within the framework of conventional superconductivity as discussed in Section IV. This certainly has strong implications on the order parameter of MgCNi$_3$.

There are evidences supporting *s*-wave superconductivity in MgCNi$_3$ from other reports. The exponential behavior of $C(T)$ has been qualitatively confirmed in [19]. Both NMR and tunneling spectroscopy revealed an *s*-wave pairing in MgCNi$_3$ [47,48]. It is noted that the results favoring unconventional pairing were reported in another tunneling spectroscopy work [35].

## B. Are there strong magnetic fluctuation effects on $T_c$ in MgCNi$_3$?

The above question comes out naturally since nominally 60% of the atoms in MgCNi$_3$ are Ni. Consequently, this has been one of the focused issues in many theoretical works. To further investigate this issue, $T_c$ can be estimated by the McMillan formula $T_c=(\hbar\check{S}_D/1.45)\exp\{-1.04(1+\lambda)/[\lambda-\mu^*(1+0.62\lambda)]\}$, where $\mu^*$ characterizes the electron-electron repulsion [49]. Taking the Fermi energy $E_F \approx 6$ eV from the energy band calculations [5,8], $\mu^*$ is estimated to be 0.15, and $T_c$=8.5 K is thus estimated by the above McMillan formula with $\lambda$=0.83. This impressing agreement with the observed $T_c$ implies that the magnetic coupling strength $\lambda_{spin}$, if it existed, would be very small. This is consistent with the conclusion reported in [17]. For comparison, $\lambda_{spin}$=0.1 would probably lower $T_c$ to 3.7 K. Should such a small $\lambda_{spin}$ have turned the order parameter into $p$-wave pairing, the physics would have been unusual. Considering only the Ni $d$ contribution would effectively make $E_F$ smaller and thus lower $T_c$, leaving possible $\lambda_{spin}$ even smaller. ($E_F$=4 eV leads to $T_c$=7.6 K which is even closer to that of the present sample.)

At least two theoretical works also suggest that the superconductivity in MgCNi$_3$ is described properly by the conventional BCS phonon mechanism [16,17]. However, a weak magnetic coupling to electrons can not be ruled out. The positive $dT_c/dP$ could be due to the weakening of the spin fluctuations with increasing $P$ [18,45]. An enhancement of the spin fluctuations with decreasing $T$ was observed in NMR experiments [47]. The x-ray photoemission spectroscopy also indicates that the electron-electron correlation may be important to physical properties of MgCNi$_3$ [50].

## C. Experiments vs. theoretical calculations

The above experimental evidences clearly demonstrate that superconductivity in MgCNi$_3$ is well explained within the conventional electron-phonon interaction scenario, at most with minor modifications from the magnetic interaction. However, there are *serious* discrepancies between experimental and band structure calculation results.

Both the transport properties of the Hall effect and the thermoelectric power effect indicate that the carriers responsible for the electrical transport are electrons [12,13,45]. The band structure calculations predict that hole states dominate at $E_F$. The pressure effects strongly suggest a positive slope of $N(E_F)$ opposite to that predicated by the theoretical calculations. Naively, these two difficulties can be resolved with the two band model mentioned above. However, this explanation implies that $T_c$ of MgCNi$_3$ is mainly ascribed to the nearly empty band with a cube shaped Fermi surface. This scenario has not been theoretically investigated in detail yet. More curiously, the photoemission spectroscopy observed the identical electronic density of states below $E_F$ for different $T_c$'s [50]. The vHs is much less sharp than predicted by calculations [48]. If the photoemission results are further confirmed and are not due the sample surface effects, there would exist stringent challenges for future theoretical studies on MgCNi$_3$.

It is of interest to compared the observed $\chi_n$ with the calculated bare band $N(E_F)$. Taking $\lambda$=0.83 into the renormalization factor (1+$\lambda$), the observed $\chi_n$ leads to $N(E_F)$=7.79 state/eV cell. This value is slightly larger than the upper calculated value ~5.5 state/eV cell in the literatures [17,19]. Nevertheless, it has been shown that carbon deficiency, which is common in MgCNi$_3$, could increase $N(E_F)$ [16]. Furthermore, the broadening of vHs due to impurity scattering is another alternative to a reconciliation. On the other hand, if the nearly empty band is responsible for the transport properties and superconductivity, it is not obvious how the band structure calculations reconcile $N(E_F)$ with the observed $\chi_n$.

## VII. Conclusions

The aim of this article is to convince the readers that MgCNi$_3$ is an interesting new superconductor. In this article, evidences are provided to show that MgCNi$_3$ is a conventional superconductor with *s*-wave order parameter. The properties related to superconductivity are consistent with each other in the context of superconductivity due to the moderate electron-phonon coupling. However, there exist major discrepancies between the results from experiments and from the band structure calculations. Further theoretical studies on the role of each band are indispensable to shed light on these puzzles. New experiments on MgCNi$_3$, especially the bulk sensitive ones or the ones with better sample quality, would be key to a full understanding of MgCNi$_3$.

## Acknowledgements

We are grateful to C.-Q. Jin for providing the samples. Especially thanks to S. Mollah for manuscript preparation. Help in experiments from C.-J. Liu, W. L. Huang, P. L. Ho, H. L. Huang is appreciated. This work was supported by National Science Council, Taiwan, Republic of China under contract Nos. NSC91-2112-M-009-046 and NSC91-2112-M-110-005.

| Estimate approaches | $H_{c2}$ (T) |
|---|---|
| $T_c$ transition by by $C$ and ... measurements; Spin-orbit interaction and spin paramagnetic effect not included. | 13.2±0.7 |
| $T_c$ transition by by $C$ and ... measurements; Spin-orbit interaction and spin paramagnetic effect included. | 11.2±0.6 |
| $d\chi/dH$ | 11.5±0.6 |

Table I.

|  | MgCNi$_3$ | Nb$_{0.5}$Ti$_{0.5}$ | Nb |
| --- | --- | --- | --- |
| $T_c$ (K) | 6.4 | 9.3 | 9.2 |
| $\Delta C/\gamma_n T_c$ | 1.97 | ~1.9 | 1.87 |
| $\ln(\Theta_D/T_c)$ | 3.79 | 3.23 | 3.40 |
| $2\Delta/kT_c$ | ≥4 | 3.9 | 3.80 |
| $H_{c2}$ (T) | 11.5 | 14.2 | ~0.2 |
| $\Theta_D$ (K) | 287 | 236 | 275 |
| $\gamma_n$ (mJ/mol K$^2$) | 33.6 (11.2/Ni) | 10.7 | 7.79 |

Table II.

**Table Captions**

TABLE I. Estimates of $H_{c2}$ in MgCNi$_3$ by different approaches.

TABLE II. Comparison between MgCNi$_3$, Nb$_{0.5}$Ti$_{0.5}$, and Nb. Parameters of MgCNi$_3$ are similar to those of Nb$_{0.5}$Ti$_{0.5}$ and Nb. $H_{c2}$ of Nb is much smaller than those of the others because Nb$_{0.5}$Ti$_{0.5}$ and MgCNi$_3$ are typical type II superconductors while Nb is nearly type II. Parameters of MgCNi$_3$ are from the present work, and those of Nb$_{0.5}$Ti$_{0.5}$ and Nb are from Refs. [51-54].

**Figure captions**

FIG. 1.  Temperature ($T$) dependence of resistivity (...) and thermoelectric power ($S$) for sample A at ambient pressure. The inset shows ... of the three samples A, B and C near $T_c$.

FIG. 2.  $C(T,H)/T$ vs. $T^2$ of MgCNi$_3$ for $H=0$ to 8 T.

FIG. 3.  (a) $\Delta C(T)/T$ vs. $T$. The data are presented as the solid circles. The solid line is the BCS $\Delta C(T)/T$ with $2\Delta/kT_c=4$. Deviation at low temperatures from the solid line is due to the magnetic contribution of a small amount of the paramagnetic centers in the sample. Inset: entropy difference $\Delta S$ by integration of $\Delta C(T)/T$ according to the data above 3 K and the solid line below 3 K. (b) The dashed lines are determined by the conservation of entropy around the anomaly to estimate $\Delta C/T_c$ at $T_c$.

FIG. 4.  $C_{es}$ of MgCNi$_3$ in the superconducting state is plotted on a logarithmic scale vs. $T_c/T$. The straight line is the fit from 2 to 4.5 K.

FIG. 5.  Magnetic field dependence of (a) $\chi$ and (b) $\delta C/T$ at $T=2$ K. The straight lines are linear fits of the data for $H\geq 4$ T implying $\delta\chi \propto H$. In (b), the fitting range is from 1 to 8 T. Data below $H=1$ T deviate from the linear behavior due to flux line interactions at low $H$. The fits by $\delta\chi(H) \propto H^{1/2}$ is shown as the dashed line in (b).

FIG. 6.  MgCNi$_3$ ...($T$) in $H$ near the transition. The applied magnetic fields from right to left are 0, 0.1, 0.2, 1, 1.5, 2, 4, 6, 8 in T, respectively.

FIG. 7.  $H_{c2}$ of MgCNi$_3$. Open and solid circles denote the results from C and ... measurements, respectively. The straight lines are the linear fits of data with $H\leq 4$ T (roughly corresponding to $0.8 \leq T/T_c < 1$). The zero field data were excluded for the fitting.

FIG. 8.  $H_{c2}$ vs. $(dH_{c2}/dT)_{T_c} T_c$ from different works. Solid circle: Ref. [34]. Open circle: Ref. [19]. Solid square: Ref. [12]. Open square: Ref. [36]. Values of $H_{c2}$: from $d\chi/dH$ in [34]; the extrapolation to $T=0$ in [19]; from $R$-$H$ in [12]; the extrapolation to $T=0$ in [36] by the present authors. None of the values is obtained from WHH formula. The solid line is from the best fit of $H_{c2} \approx aT_c(dH_{c2}/dT)_{T_c}$.

FIG.9.    Pressure (*P*) dependence of $T_c$ of samples A, B and C.

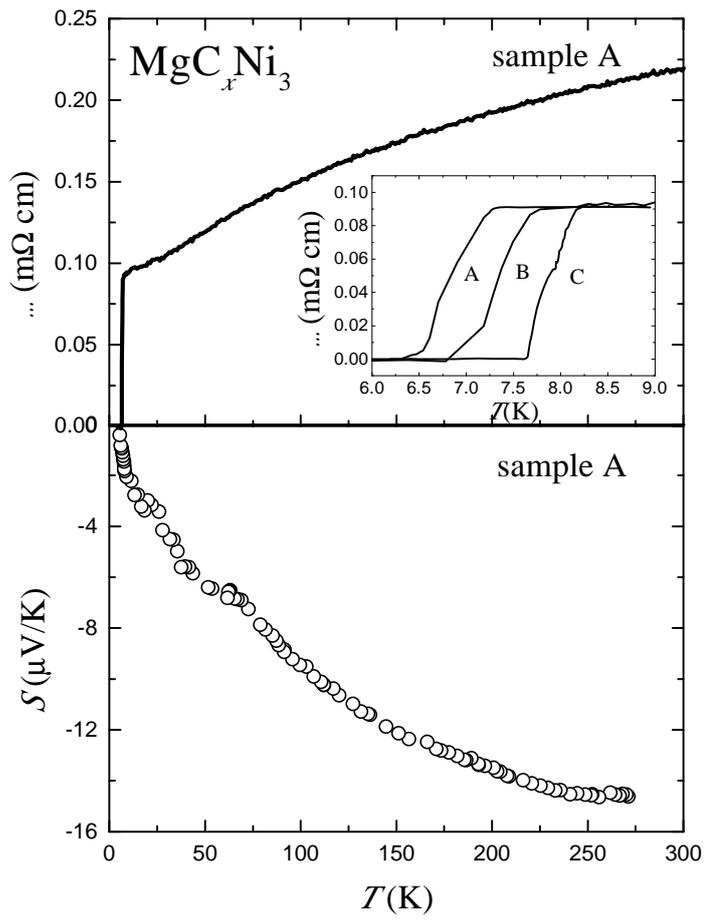

Lin and Yang Fig. 1

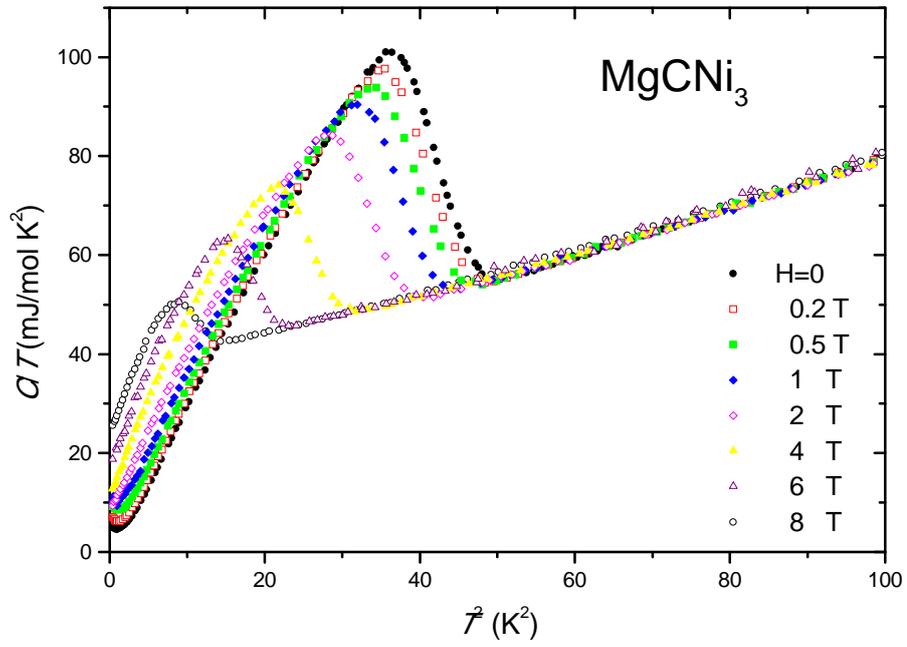

Lin and Yang Fig. 2

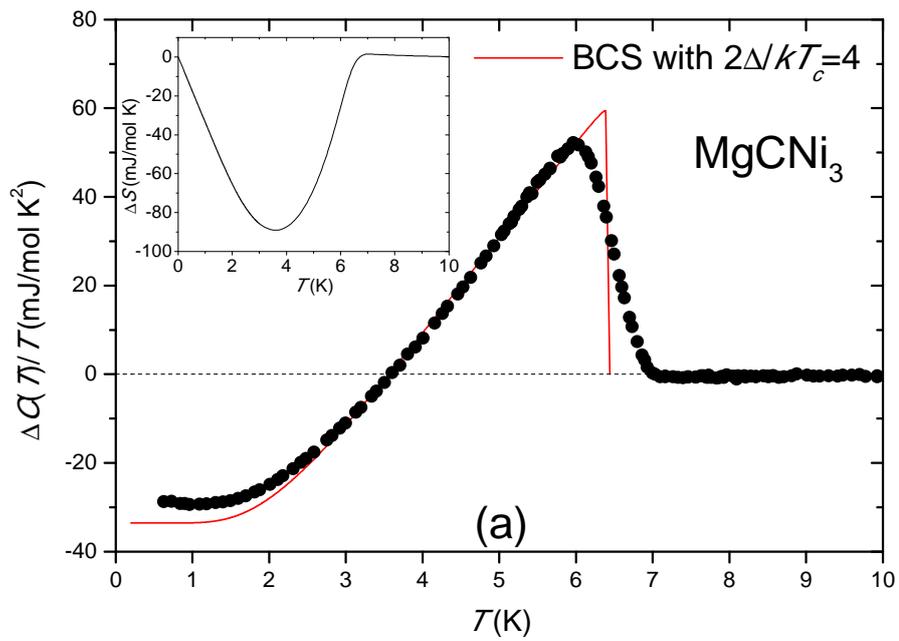

Lin and Yang Fig. 3(a)

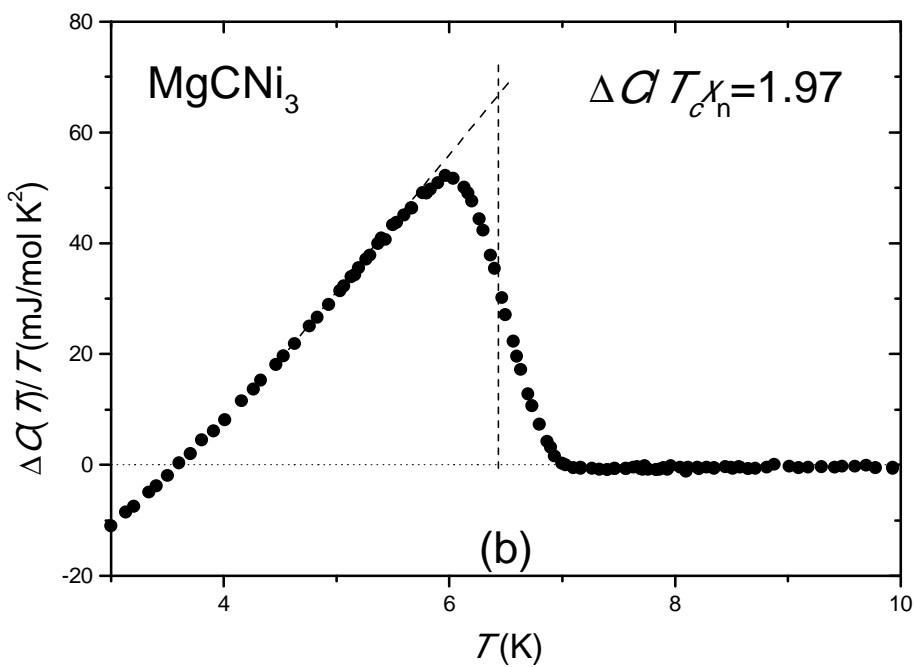

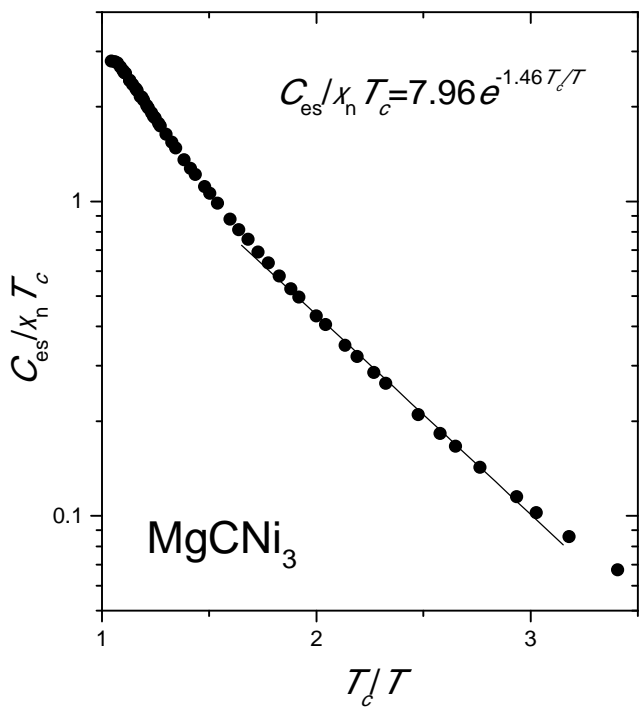

Lin and Yang Fig. 4

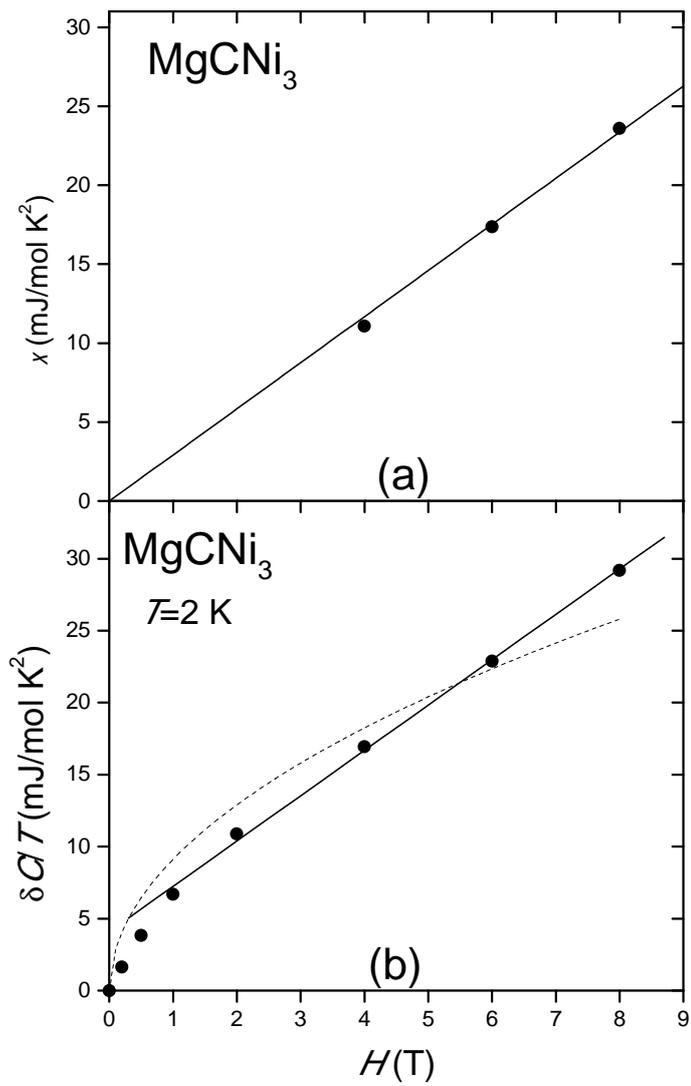

Fig. 5 Lin and Yang

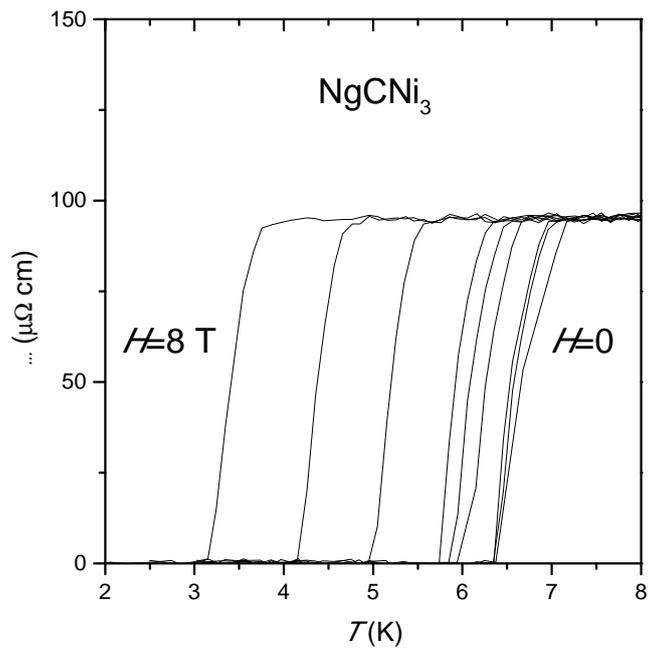

Fig. 6. Lin and Yang

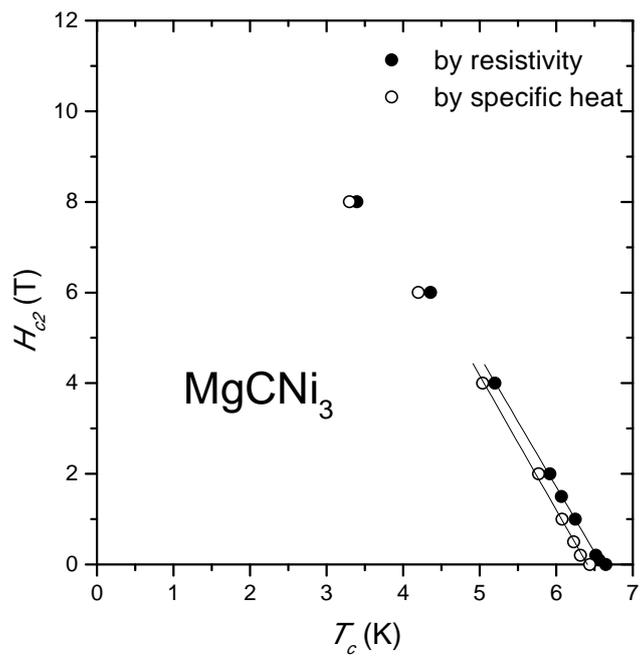

Fig. 7. Lin and Yang

Lin and Yang Fig. 8

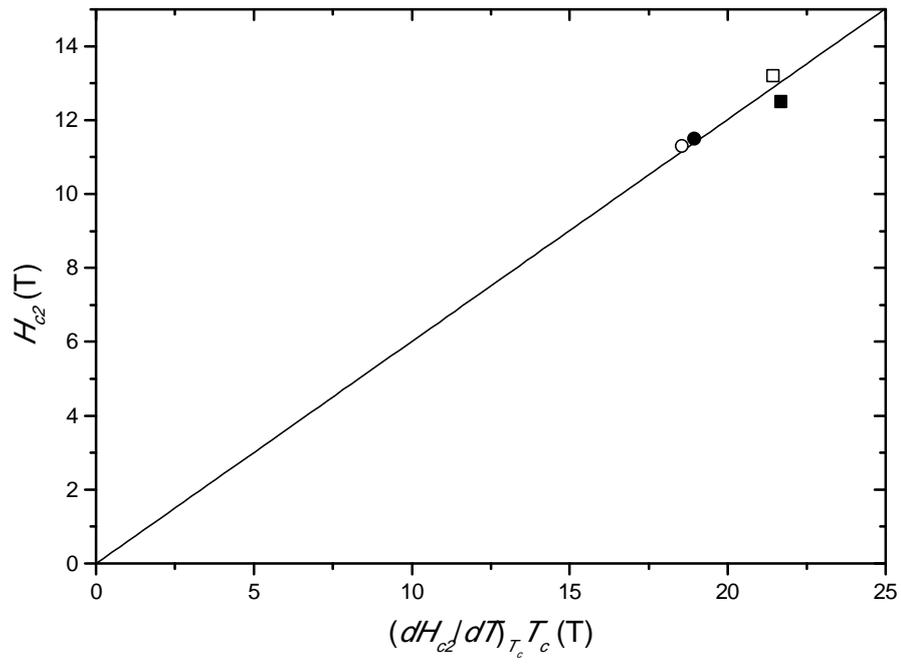

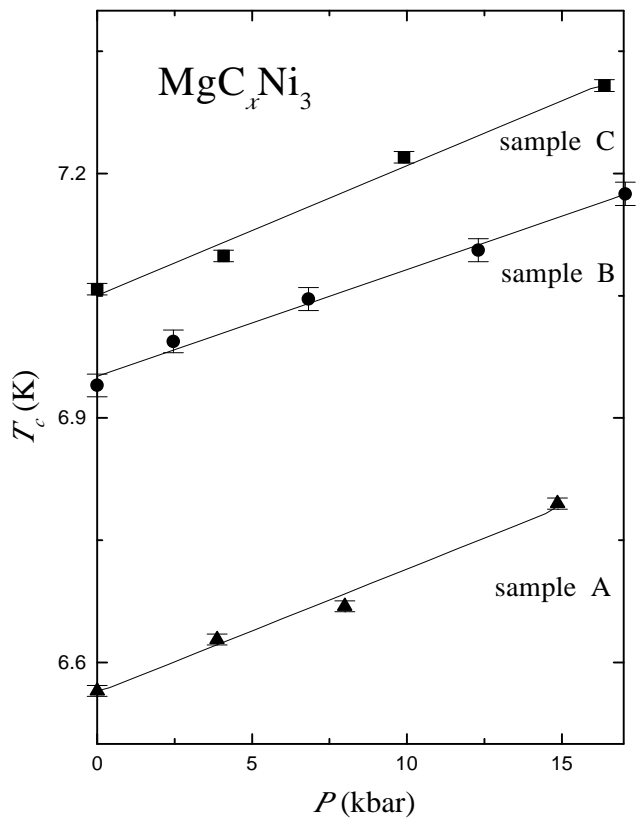

Lin and Yang Fig. 9